\documentclass[aps,prl,twocolumn,10pt]{revtex4-1}

\usepackage{ifthen}
\usepackage{hyperref}
\usepackage{centernot}

\newcommand{\psec}[1]{\emph{#1.}---}

\newcommand{\bq}{\begin{equation}}
\newcommand{\eq}{\end{equation}}
\newcommand{\bqa}{\begin{eqnarray}}
\newcommand{\eqa}{\end{eqnarray}}

\newcommand{\rmd}{{\rm d}}
\newcommand{\Om}{\Omega_{\rm m}}
\newcommand{\rhom}{\rho_{\rm m}}
\newcommand{\Deltam}{\Delta_{\rm m}}
\newcommand{\aK}{\alpha_{\rm K}}
\newcommand{\aM}{\alpha_{\rm M}}
\newcommand{\aB}{\alpha_{\rm B}}
\newcommand{\aT}{\alpha_{\rm T}}
\newcommand{\cs}{c_{\rm s}}
\newcommand{\cT}{c_{\rm T}}

\begin{document}

\title{Challenges to Self-Acceleration in Modified Gravity from Gravitational Waves and Large-Scale Structure}

\author{Lucas~Lombriser}
\affiliation{Institute for Astronomy, University of Edinburgh, Royal Observatory, Blackford Hill, Edinburgh, EH9~3HJ, U.K.}
\author{Nelson~A.~Lima}
\affiliation{Institute for Astronomy, University of Edinburgh, Royal Observatory, Blackford Hill, Edinburgh, EH9~3HJ, U.K.}

\date{\today}


\begin{abstract}
With the advent of gravitational-wave astronomy marked by the aLIGO GW150914 and GW151226 observations, a measurement of the cosmological speed of gravity will likely soon be realized.
We show that a confirmation of equality to the speed of light as indicated by indirect Galactic observations will have important consequences for a very large class of alternative explanations of the late-time accelerated expansion of our Universe.
It will break the dark degeneracy of self-accelerated Horndeski scalar-tensor theories in the large-scale structure that currently limits a rigorous discrimination between acceleration from modified gravity and from a cosmological constant or dark energy.
Signatures of a self-acceleration must then manifest in the linear, unscreened cosmological structure.
We describe the minimal modification required for self-acceleration with standard gravitational-wave speed and show that its maximum likelihood yields a $3\sigma$ poorer fit to cosmological observations compared to a cosmological constant.
Hence, equality between the speeds challenges the concept of cosmic acceleration from a genuine scalar-tensor modification of gravity.
\end{abstract}


\maketitle


\psec{Introduction}
Nearly two decades after the discovery of the late-time accelerated expansion of our Universe~\cite{riess:98,perlmutter:98}, unraveling the physical nature underlying the effect remains a difficult puzzle and a prime endeavor in cosmology.
In the concordance $\Lambda$ Cold Dark Matter ($\Lambda$CDM) model, a cosmological constant $\Lambda$ contributes the bulk of the energy density in the present cosmos and accelerates its expansion in accordance with Einstein's Theory of General Relativity (GR).
While $\Lambda$ may be attributed to a vacuum energy, its observed value is inexplicably small to theory.
Alternatively, it has been conjectured that a modification of gravity at cosmological scales may be responsible for the effect (see~\cite{clifton:11,koyama:15,joyce:16} for reviews).
However, stringent constraints from the verification of GR in the Solar System must be satisfied and a few screening mechanisms have been identified that can restore GR in high-density regions while still permitting significant modifications at low densities at cosmological scales (see~\cite{joyce:14} for a review).
The simplest infrared remnant we can conceptualize to arise from a potentially more fundamental theory of gravity is a single effective scalar degree of freedom that permeates the universe and may modify gravity to drive cosmic acceleration.
The most general scalar-tensor theories have been shown to contain enough freedom to recover the cosmological background expansion and large-scale structure of $\Lambda$CDM without the need of a cosmological constant to explain cosmic acceleration, while remaining theoretically consistent~\cite{lombriser:14b,lombriser:15c}.
However, the requirement of self-acceleration for degenerate models implies a symptomatic and detectable deviation of the propagation speed of gravitational waves from the speed of light~\cite{lombriser:15c}.

In contrast, the Hulse-Taylor binary system and observations of ultra-high energy cosmic rays set tight constraints on possible deviations in our Galactic environment, which is a strong indication that the two speeds should also agree at cosmological scales~\cite{moore:01,jimenez:15,lombriser:15c}.
With the recent breakthrough in the direct detection of gravitational waves with the Advanced Laser Interferometer Gravitational-Wave Observatory (aLIGO)~\cite{GW150914,GW151226}, a cosmological measurement will likely soon be realized. It will provide the crucial constraint to break this dark degeneracy.
Hence, provided the confirmation of a standard speed of gravity at cosmological scales, a genuine self-acceleration generated by a scalar-tensor modification of gravity must manifest in the large-scale structure.
In this Rapid Communication, we determine the minimal cosmological signatures such a scenario must produce which cannot be suppressed by screening mechanisms.
We then examine whether these conservative signatures are compatible with observations.


\psec{Self-acceleration from modifying gravity}
Conceptually, modified gravity may be thought of as the presence of a force altering the motion of a freely falling particle along a geodesic which is determined from a metric that obeys the Einstein field equations but where the matter sector is modified.
Alternatively to this Einstein frame, one can define a Jordan frame, where the motion of the particle is still governed by the geodesic but the metric satisfies a modified field equation with a conventional energy-momentum tensor.
While in GR there is no difference between the two frames, when gravity is modified the respective equations of motion are distinct in form.
Note however that for a general modification defined in Jordan frame, an Einstein frame does not always exist for all metrics.
We shall therefore more generally define an \emph{Einstein--Friedmann~frame}~\cite{lombriser:15c} which only requires that the cosmological background evolution equations, which are the relevant equations to address cosmic acceleration, can be mapped into their standard form.

We assume a four-dimensional, spatially statistically homogeneous and isotropic as well as flat ($k_0=0$) universe with the Jordan-frame Friedmann-Lema\^itre-Robertson-Walker (FLRW) metric given by the line element $\rmd s^2 = -\rmd t^2 + a^2(t) \rmd {\bf x}^2$, where we set the speed of light in vacuum to unity.
The scale factor $a(t)$ determines the Hubble parameter $H(t)\equiv\rmd\ln a/\rmd t$.
We then specify the conformal factor $\Omega$ mapping the Jordan-frame metric $g_{\mu\nu}$ to the metric $\tilde{g}_{\mu\nu} =  \Omega \, g_{\mu\nu}$ in the Einstein-Friedmann frame and transform the time coordinate to cast $\tilde{g}_{\mu\nu}$ in the same form as the Jordan-frame FLRW metric with $\tilde{a}(\tilde{t})$ denoting the scale factor as function of proper time $\tilde{t}$ in the Einstein-Friedmann frame.
For the cosmic acceleration observed in the Jordan frame to be genuinely attributed to a gravitational modification rather than the contribution of a cosmological constant or a dark energy in the matter sector, there should conservatively be no positive acceleration in $\tilde{a}(\tilde{t})$ (see, e.g., Ref.~\cite{wang:12}).
Hence, whereas the observed late-time accelerated expansion implies
\bq
 \frac{\rmd^2 a}{\rmd t^2} = a\,H^2\left(1+\frac{H'}{H}\right) > 0 \label{eq:accjordan}
\eq
for $a\gtrsim0.6$, a genuine self-acceleration from a generic modification of gravity must satisfy~\cite{lombriser:15c}
\bq
 \frac{\rmd^2\tilde{a}}{\rmd \tilde{t}^2} = \frac{a\,H^2}{\sqrt{\Omega}} \left[\left(1+\frac{H'}{H}\right)\left(1+\frac{1}{2}\frac{\Omega'}{\Omega}\right) + \frac{1}{2}\left(\frac{\Omega'}{\Omega}\right)' \right] \leq 0 \label{eq:acceinstein}
\eq
with primes indicating derivatives with respect to $\ln a$.
Thus, self-acceleration implies $-\Omega'/\Omega\gtrsim\mathcal{O}(10^{-1})$.

Naturally, the gravitational modification should also have an impact on structure formation and the propagation of gravitational waves.
To describe how large-scale structure is affected, we consider linear scalar perturbations of the FLRW metric, adopting the Newtonian gauge with $\Psi\equiv\delta g_{00}/(2g_{00})$ and $\Phi\equiv\delta g_{ii}/(2g_{ii})$, working in Fourier space,
and restricting to a universe dominated by pressureless dust $p_{\rm m}=0$ with background matter density $\rhom$ and perturbation $\Deltam$ in total matter gauge.
We then characterize the modified Einstein field equations by an effective modification of the Poisson equation, $\mu$, and a gravitational slip, $\gamma$,~\cite{uzan:06,caldwell:07,zhang:07}
\bq
 k_H^2\Psi = -\frac{\kappa^2\rhom}{2H^2}\mu(a,k)\Deltam \,, \ \ \ \ \
 \Phi = -\gamma(a,k)\Psi \,, \label{eq:mugamma}
\eq
respectively, where $k_H\equiv k/(aH)$ and $\kappa^2\equiv8\pi\,G$ with bare gravitational constant $G$.
The standard energy-momentum conservation equations then close the system of constraint and differential equations determining the evolution of the scalar modes.
The modified cosmological propagation of tensor modes is described by the linear traceless spatial metric perturbation $h_{ij}\equiv g_{ij}/g_{ii}$ with~\cite{saltas:14}
\bq
 h_{ij}'' + \left(\nu + 2 + \frac{H'}{H} \right)h_{ij}' + c_{\rm T}^2k_H^2 h_{ij} = 0 \,, \label{eq:gw}
\eq
where $\nu$ describes a running of the gravitational coupling or Planck mass, altering the damping term, and $c_{\rm T}$ denotes the tensor sound speed, which can differ from the speed of light.
GR is recovered in the limit of $\mu=\gamma=\nu=c_{\rm T}=1$.


\psec{Effective field theory}
As an alternative to cosmic acceleration from a cosmological constant $\Lambda$, we shall consider the presence of a single low-energy effective scalar field that permeates our Universe and couples non-minimally to the metric, modifying gravity such to cause the late-time expansion to accelerate.
Horndeski gravity~\cite{horndeski:74} describes the most general four-dimensional, local, and Lorentz-covariant scalar-tensor theory with at most second-order derivatives of the scalar and tensor fields in the Euler-Lagrange equations.
It embeds a large fraction of the modified gravity and dark energy models that have been proposed as alternative explanations for cosmic acceleration~\cite{clifton:11,koyama:15,joyce:16}.
To conveniently describe its cosmological background and linear perturbations, we adopt the effective field theory (EFT) of cosmic acceleration (see Ref.~\cite{gleyzes:14b} for a review). 
The formalism disentangles the Hubble parameter $H$ as a generally freely time-dependent function that determines the cosmological background evolution.
Linear perturbations are then characterized by an additional four free functions of time, each describing a different physical property of the scalar-tensor theory: $\aK$, $\aM$, $\aB$, and $\aT$.
The kineticity $\aK$ parametrizes a kinetic energy contribution from the scalar field that can cause it to cluster at ultra-large scales.
The running of the gravitational coupling with Planck mass $M^2$ at the rate $\aM\equiv(\ln M^2)'$ introduces a gravitational slip between $\Psi$ and $\Phi$ and modifies the damping of tensor waves.
The parameter $\aB$ describes interactions of the scalar and metric fields through braiding, or mixing, of their kinetic contributions, giving rise to clustering of the scalar field on small scales.
Finally, $\aT$ parametrizes a departure of the speed of tensor modes from the speed of light, also contributing to the gravitational slip.
$\Lambda$CDM is recovered when $\alpha_i=0$ $\forall i$.

The linearly perturbed modified Einstein equations in EFT~\cite{bellini:14,gleyzes:14b,lombriser:15b} at the observable sub-Hubble scales ($k_H\gg1$) can be described by Eqs.~(\ref{eq:mugamma}) with
\bqa
 \mu_{\infty} & = & \frac{2\left[\aB(1+\aT)-\aM+\aT\right]^2 + \alpha(1+\aT) c_{\rm s}^2}{\alpha \cs^2 \kappa^2M^2} \,, \label{eq:muinfH} \\
 \gamma_{\infty} & = & \frac{2\aB \left[\aB(1+\aT)-\aM+\aT\right] + \alpha c_{\rm s}^2}{2\left[\aB(1+\aT)-\aM+\aT\right]^2 + \alpha (1+\aT) \cs^2} \,, \label{eq:gammainfH}
\eqa
where $\alpha \equiv 6 \aB^2 + \aK$ and $\cs^2$ denotes the sound speed of the scalar mode.
Note that $\cs^2$ depends on $\alpha_i$ and $H$, whereby $\alpha\,\cs^2$, and thus $\mu_{\infty}$ and $\gamma_{\infty}$, are independent of $\aK$.
For $\cs^2\centernot\ll1$, the scale dependence in $\mu(a,k)$ and $\gamma(a,k)$ for natural self-accelerated models is governed by a Compton wavelength of order the Hubble scale~\cite{wang:12}. Hence, scale independence in $\mu\simeq\mu_{\infty}(a)$ and $\gamma\simeq\gamma_{\infty}(a)$ serves as a good approximation for our observational purpose.
For the tensor perturbations in Eq.~(\ref{eq:gw}) it follows that $\nu=1+\aM$ and $\cT^2=1+\aT$~\cite{gleyzes:14b,saltas:14}, directly relating the modifications of the scalar and tenor modes.

Importantly, following Eqs.~(\ref{eq:accjordan}) and (\ref{eq:acceinstein}), a cosmic acceleration which can genuinely be attributed to a modification of gravity requires~\cite{lombriser:15c}
\bq
 -\frac{\Omega'}{\Omega} = -\left(\aM + \frac{\aT'}{1+\aT}\right) \gtrsim \mathcal{O}(10^{-1}) \label{eq:nonzeroaMaT}
\eq
at late times.
This implies that at least one of either $\aM$ or $\aT'$ should be nonzero.


\psec{Implication of $\aT\simeq0$}
Importantly, self-accelerated Horndeski scalar-tensor gravity contains enough freedom to reproduce the $\Lambda$CDM values $\mu(a,k)=\gamma(a,k)=1$ in Eqs.~(\ref{eq:mugamma}) at all scales and additionally allow a recovery of the concordance model background expansion history~\cite{lombriser:14b,lombriser:15c}.
Thus, cosmic acceleration from a cosmological constant or a scalar-tensor modification of gravity cannot rigorously be discriminated using geometric probes and the large-scale structure only.
The capacity for such a degeneracy or linear shielding effect~\cite{lombriser:14b}, however, requires that $\mu=\gamma=1$ in Eqs.~(\ref{eq:muinfH}) and (\ref{eq:gammainfH}) and with recovery of
$H=H_{\Lambda{\rm CDM}}$, this yields a nonlinear differential equation relating $\aM$ to $\aB$ and the condition~\cite{lombriser:15c}
\bq
 \aT = \frac{\kappa^2M^2-1}{(1+\aB)\kappa^2M^2-1}\aM \,. \label{eq:aT}
\eq
Therefore for a degeneracy to occur and for a self-acceleration with $\Omega'\neq0$, we must have non-vanishing $\aM$, $\aB$ and in particular $\aT$, which generally implies a deviation of the propagation speed of gravitational waves $\cT$ from the speed of light.

Note that there already are strong indications that $\aT\simeq0$.
For instance, the observation of ultra-high energy cosmic rays implies that subluminal deviations from $\cT=1$ are constrained to $\mathcal{O}(10^{-19}-10^{-15})$ as otherwise they would decay due to gravitational Cherenkov radiation before reaching Earth~\cite{moore:01}.
Furthermore, the Hulse-Taylor binary system places percent-level constraints on deviations from $\cT=1$~\cite{jimenez:15}.
It can, however, currently not fully be ruled out that a screening mechanism may not suppress the cosmological deviations in $\cT$ in the Galactic environment from which these constraints are inferred.
However, this is unlikely given the high degree of suppression required and, moreover, it was shown in Ref.~\cite{jimenez:15} that the screening mechanism conventionally associated with the modifications introducing $\aT\neq0$ is not capable of such a suppression.

To be fully conservative, however, we ultimately require a cosmological constraint on $\cT$.
The recent aLIGO observation of a gravitational wave (GW150914)~\cite{GW150914} emitted by a merger of two black holes at $(0.4\pm0.2)$~Gpc and the nearly simultaneous detection of a weak short gamma-ray burst in the Fermi Gamma-ray Burst Monitor~\cite{GRB150914}, which has been claimed to be connected with the same event, would constrain $\aT\simeq0$ for the purpose of self-acceleration in Eq.~(\ref{eq:nonzeroaMaT}).
While the association of the two measurements is controversial~\cite{savchenko:16,abbott:16}, it demonstrates that with the advent of direct gravitational wave observations a cosmological measurement of $\cT$ will likely soon be realized, potentially already in the second aLIGO observing run.
A confirmation of $\cT\simeq1$ would then imply that self-acceleration must solely be due to an evolution of the gravitational coupling with $-\aM\simeq-\Omega'/\Omega\gtrsim\mathcal{O}(10^{-1})$.
Furthermore, from Eq.~(\ref{eq:aT}) it then follows that there cannot be any degeneracy effect and that the self-accelerated modification must manifest in the large-scale structure.


\psec{Minimal self-accelerated modification}
In the following, we adopt $\alpha_{\rm T}=0$ and determine the minimal signature a self-accelerated modification must leave in the large-scale structure.
With $\Omega'/\Omega = \alpha_{\rm M}$ in Eq.~(\ref{eq:acceinstein}), we find that self-acceleration must satisfy
\begin{equation}
 \left( 1 + \frac{H'}{H} \right) \left( 1 +\frac{1}{2} \alpha_{\rm M} \right) + \frac{1}{2}\alpha_{\rm M}' \leq 0 \,, \label{eq:einsteinacc}
\end{equation}
where positive acceleration in Jordan frame implies $(1+H'/H)>0$ from Eq.~(\ref{eq:accjordan}).
We define $a_{\rm acc}$ as the scale factor at which this term vanishes.
For a $\Lambda$CDM expansion history this implies $a_{\rm acc} = [\Omega_m/(1-\Omega_m)/2]^{1/3}$.
Since we want to recover $\Lambda$CDM at early times, we set $\alpha_i=0$ $\forall i$ for $a\leq a_{\rm acc}$ and also adopt $H=H_{\Lambda{\rm CDM}}$ to render the models indistinguishable at the background level, restricting to the strictly necessary modifications in the linear fluctuations in the accelerated era $a>a_{\rm acc}$.
Integration of Eq.~(\ref{eq:einsteinacc}) yields $\alpha_{\rm M} \leq C/aH - 2$, where $C$ is an integration constant, which from requiring that the modification vanishes at $a=a_{\rm acc}$ we set to $C = 2 H_0 a_{\rm acc} \sqrt{3(1-\Omega_m)}$.
We have $\alpha_{\rm M}(a>a_{\rm acc}) < \alpha_{\rm M}(a\leq a_{\rm acc}) = 0$.
Another integration yields $M^2$ and with normalization at $a_{\rm acc}$,
\begin{equation}
 \kappa^2M^2 \leq \left(\frac{a_{\rm acc}}{a}\right)^2 e^{C(\chi_{\rm acc}-\chi)} \,,
\end{equation}
where $\chi$ denotes the comoving distance.
The modification is minimal when we have equality.
To avoid ghost and gradient instabilities of scalar and tensor fluctuations, the modified models must satisfy $M^2>0$, $\alpha>0$, and $\cs^2>0$~\cite{bellini:14}.
As we want to recover GR at early times and require self-acceleration, we further have $\kappa^2M^2\leq1$.
From Eqs.~(\ref{eq:mugamma}) and (\ref{eq:muinfH}), this implies that the modification must yield an enhancement of gravity with $\mu\geq1$.
To minimize the effect on the growth of structure with respect to GR, we require $\alpha_{\rm B} = \alpha_{\rm M}$ and hence, we obtain $\mu = (\kappa^2M^2)^{-1}$ and the GR value $\gamma = 1$.
This fully specifies the minimal Horndeski scalar-tensor modification required for a genuine self-acceleration and yields $\mu(a\leq a_{\rm acc}\simeq0.6)=1$ increasing to $\mu(a=1)\simeq1.04$ today with $\gamma=1$ at any scale factor.


\psec{Challenges from large-scale structure}
We conduct a parameter estimation analysis of the minimal self-accelerated modification and $\Lambda$CDM employing {\sc cosmomc}~\cite{lewis:02} and {\sc mgcamb}~\cite{hojjati:11}, where we adopt the default basic cosmological parameters~\cite{planck:15} for variation.
Importantly, note that the modified model does not introduce any new parameters.
We use geometric probes from supernovae~\cite{betoule:14}, baryon acoustic oscillation distance~\cite{beutler:11,padmanabhan:12,anderson:12}, and local $H_0$~\cite{efstathiou:13} measurements
as well as the Planck 2015~\cite{planck:15} cosmic microwave background (CMB) temperature fluctuation, polarization, and lensing data, including cross correlations and low-multipole measurements.
While the geometric probes only constrain background parameters, the secondary anisotropies in the CMB are sensitive to the late-time modifications in $\mu$.
In addition, we use the $E_G\equiv\Sigma\,\Om/(\Deltam'/\Deltam)$~\cite{zhang:07} measurement of Ref.~\cite{reyes:10}, testing a combination of weak gravitational lensing, galaxy clustering, and structure growth rate, where $\Sigma=(1+\gamma)\mu/2$ describes the modified Poisson equation of the lensing potential $\Phi_-\equiv(\Phi-\Psi)/2$.
Note that while there are newer measurements of $E_G$, the low-redshift constraint of Ref.~\cite{reyes:10} is most suitable for our purpose ($a>a_{\rm acc}$).
Finally, we use the publicly available data~\cite{ho:08} of cross correlations of the CMB temperature field with foreground galaxies through the integrated Sachs-Wolfe (ISW) effect, where we follow Refs.~\cite{lombriser:09,lombriser:10} in the implementation of the gravitational modifications in the likelihood code.

We find that in comparison to a cosmological constant and given $\cT\simeq1$, the minimal self-accelerated modification produces a $\Delta\chi^2=9.2$~($3\sigma$) poorer maximum likelihood fit to the combination of these data sets.
For an approximation of the Bayes factor using the average likelihood over the chain samples~\cite{lewis:02}, we find $B\approx\langle\mathcal{L}_{\Lambda}\rangle_{\rm s}/\langle\mathcal{L}_{\rm MG}\rangle_{\rm s}\simeq39$, which points towards a very strong evidence for $\Lambda$.
The difference is driven by the galaxy-ISW cross correlations, which provide a direct probe of the evolution rate of the Planck mass $\aM$ responsible for self-acceleration.
Note that the ISW effect is sensitive to $\Sigma'=-\aM\Sigma$ which modifies the change in energy of CMB photons when traversing the evolving potential $\Phi_-'\neq0$ at late times.


\psec{Caveats}
The concordance model provides a good fit to cosmological observations, which motivated our consideration of the minimal self-accelerated modification.
While we found that the fit worsened for the small deviation, we cannot strictly exclude the existence of a radical and very different modification that may improve the fit.
It is important to note, however, that due to the condition that $\aT\simeq0$ and $\mu\geq1$, a self-accelerated model cannot account for the slight preference of a weakening of gravity in the cosmological data reported, for instance, in Ref.~\cite{planck:15demg}.
On the contrary, the enhancement of gravity due to self-acceleration increases tensions in this data.
Instead of minimizing the modification in $\mu$, one may also chose to minimize it in $\Sigma$.
This can, however, not be done without knowledge of $\aB$.
Adopting the same time evolution in $\aB$ as in $\aM$, we find that this can only reduce $\Sigma$ at a few percent over $\Sigma=(\kappa^2M^2)^{-1}$ following a minimization of $\mu$, while this also increases deviations in $\mu$ and~$\gamma$.
One may also imagine that a scale dependence in $\mu$ and $\gamma$ within the Hubble horizon, arising for instance from $\cs\ll1$ may weaken the modification observed at very large scales.
However, a semi-dynamical approximation accessing these scales shows that for $\aT\simeq0$, the evolution of $\Phi$ in this limit is governed by $H$ and $\aM$ and hence directly modified by the running of the gravitational coupling~\cite{lombriser:15b}.


\psec{Discussion}
The recent observational breakthrough in the direct detection of gravitational waves will likely soon enable a measurement of the cosmological speed of gravity.
A confirmation of equality to the speed of light, strongly supported by Galactic measurements, would have important implications for self-accelerated Horndeski scalar-tensor theories.
We have shown that signatures of a self-acceleration must then manifest in the linear, unscreened cosmological structure, as the measurement of a tensor sound speed consistent with the speed of light prevents a dark degeneracy in the large-scale structure with a cosmological constant or dark energy.

We have derived the minimal modification required for a genuine self-acceleration in this scenario and found that in comparison to a cosmological constant, it produces a $3\sigma$ poorer maximum likelihood fit to cosmological observations, including geometric probes, CMB data, and constraints from weak lensing, galaxy clustering, and the structure growth rate.
But in particular, we have used galaxy-ISW cross correlations that are very sensitive to the evolving gravitational coupling responsible for a self-acceleration in a modified gravity model consistent with a gravitational wave propagation at the speed of light.
Future weak lensing measurements of the evolution of $\Phi_-$ and standard sirens~\cite{lombriser:15c}, both sensitive to $\aM$, may improve the discrimination between the models.
Although marginally still possible, our result sets a challenge to the concept of cosmic acceleration from a genuine scalar-tensor modification of gravity.


\begin{acknowledgments}
The authors thank Andrew Liddle and Andy Taylor for useful discussions and comments on the manuscript.
L.L.~was supported by a SNSF Advanced Postdoc.Mobility Fellowship (No.~161058) and the STFC Consolidated Grant for Astronomy and Astrophysics at the University of Edinburgh.
N.A.L.~acknowledges support from a FCT grant (No.~SFRH/BD/85164/2012).
Numerical computations were conducted on the COSMOS Shared Memory system at DAMTP, University of Cambridge operated on behalf of the STFC DiRAC HPC Facility. This equipment is funded by the BIS National E-infrastructure capital grant ST/J005673/1 and STFC grants ST/H008586/1, ST/K00333X/1.
Please contact the authors for access to research materials.
\end{acknowledgments}


\vfill
\bibliographystyle{arxiv_physrev_mod.bst}
\bibliography{accmg.bib}

\end{document}